\documentclass{PoS}

\usepackage{amsmath}

\title{Pion physics in two flavor strong coupling lattice QED}

\ShortTitle{Pion physics in two flavor strong coupling lattice QED}

\author{\speaker{D. J. Cecile}\\
        Box 90305 Duke University, Durham, NC 27708, USA\\
        E-mail: \email{djcecile@phy.duke.edu}}

\abstract{We consider the lattice field theory involving two flavors 
of staggered quarks which interact with $U_A(1)$ gauge fields in the 
strong coupling limit.  For massless quarks, this theory has an 
$SU_L(2)\times SU_R(2) \times U_A(1)$ symmetry.  We show explicitly how pions 
emerge through the phenomena of confinement in this theory.  We also 
show how one can incorporate the physics of the anomaly in this theory. 
Thus, our approach is a good pedagogical tool to explain how pions arise 
in real QCD.  Another advantage of our approach is that we can easily 
design efficient cluster algorithms to compute a variety of quantities 
close to the chiral limit, thus allowing us to understand the low
energy physics in a QCD-like setting from first principles.
}

\FullConference{XXIVth International Symposium on Lattice Field Theory\\
                July 23-28, 2006\\
                Tucson, Arizona, USA}

\begin{document}

\section{Introduction}


One of the outstanding problems in lattice QCD is to compute low energy 
hadronic observables, which are dominated by the physics of light quarks, 
with controlled errors. Although progress is being made in 
terms of fermion algorithms for lattice QCD, it will still be difficult to
approach realistic quark masses in the near future. As a result, 
calculations are performed at unphysically large quark masses and then 
extrapolations to realistic quark masses are performed using chiral 
perturbation theory. However, for such an approach 
to be reliable, we must know the range over which the chiral 
expansion is valid. Unfortunately, this is not very well understood. In fact a 
systematic study of chiral perturbation theory as an effective field 
theory that describes a more fundamental lattice field theory has not 
been attempted in many situations. Thus, it is useful to study a 
simpler lattice field theory with the same symmetries of lattice QCD 
and make connections with chiral perturbation theory. This is the main
motivation for our work.

We study the physics of strongly coupled lattice QED with 
two flavors of staggered fermions which we refer to as {\em our model} throughout
this manuscript. Although this model does not describe full QCD, 
it is an interesting model which has the same symmetries of two flavor QCD
and was recently used to study the chiral phase transition \cite{AM06}. 
Note that even a $U(1)$ gauge theory exhibits confinement in the strong 
coupling limit, and the taste symmetry is irrelevant since it
is maximally broken at strong couplings. Our model allows us to 
study the pion physics from a fundamental lattice field theory, very
similar to QCD, so that we can understand the usefulness of chiral 
perturbation theory as an effective description of low energy physics.
It is important to note that unless we find a way to fine tune our
model we will be dominated by lattice artifacts since at strong coupling
the pion decay constant $F_\pi$ is naturally close to the cutoff.
In order to circumvent this problem, we define our model in $d+1$
dimensions where $d=4$ is the space time dimensions. The extra dimension
plays the role of a fictitious temperature which allows us to tune to
a critical point where $F_\pi$ will be much smaller than the cutoff.
Thus, we can still explore the physics of a continuum limit even in the
strong coupling limit.

The motivation for studying a strongly coupled theory is that the gauge 
dependent degrees of freedom can be integrated over, which significantly 
simplifies the theory. Further, a new class of algorithms,  
{\em the Directed Path Algorithm}, has recently been designed for strongly 
coupled gauge theories, by which it is possible to study the 
chiral limit very efficiently \cite{DA03}. We have extended
this algorithm to our model and are currently testing it.  Here we report 
on current progress.

We also wish to compute via L\"{u}scher's method \cite{ML91}
quantities that are measurable in scattering experiments.  
Specifically, we would like to measure scattering lengths, characterize 
and understand resonances which may exist, and provide a setting 
to understand non-perturbative features that can arise in a QCD like 
theory and that go beyond the scope of chiral perturbation theory. Such a 
study appears impossible with the current technology of lattice QCD since 
some physical processes which occur for sufficiently light quarks, e.g. the 
decay of the $\rho$ meson into two pions, is forbidden at the quark masses 
currently used.

\section{Model and Symmetries}

The Euclidean space action of the $N_f=2$ QED model we consider is given by
(Note that the usual factors $\frac{1}{2}$ are absorbed into the field 
definitions):
\begin{equation}
S = - \sum_{x,\mu}
\eta_{\mu,x}\bigg[\mathrm{e}^{i\phi_{\mu,x}}{\overline\psi}_x
{\psi}_{x+\hat\mu}
-\mathrm{e}^{-i\phi_{\mu,x}}{\overline\psi}_{x+\hat\mu}{\psi}_x\bigg]
-\sum_x \bigg[m{\overline\psi}_x{\psi}_x
+\frac{\tilde c}{2}\bigg({\overline\psi}_x{\psi}_x \bigg)^2\bigg]
\label{eq1}
\end{equation}
where $x$ denotes a lattice site on a $d+1$ dimensional hypercubic lattice
$L_t \times L^d$.  $\overline\psi_x$ and $\psi_x$ are two component 
Grassman fields that represent the two quark flavors of mass $m$, 
and $\phi_{\mu,x}$ is the $U(1)$ gauge field through which the fields 
interact.  Note that $\mu$ runs over the temporal and spatial directions 
$0,1,2,...,d$ with $0$ denoting the temporal direction. The usual staggered 
fermion phase factors $\eta_{\mu,x}$ obey the relations: $\eta_{0,x}^2 = T$ 
and $\eta_{i,x}^2 = 1$ for $i=1,2,...,d$. The parameter $T$ is the fictitious
temperature which will be used to the continuum limit. The coupling $\tilde c$ 
will set the strength of the anomaly.

We now discuss how our model has the same symmetries and symmetry breaking
patterns of full QCD. It is first useful to note that any sum over lattice 
sites can be decomposed into a sum over {\it even} and {\it odd} sites.  
At $\tilde c,m = 0$, the action exhibits a global 
$SU(2)\times SU(2)\times U_A(1)$ symmetry.  In particular, the action is 
invariant under the following $U_A(1)$ 
and $SU_L(2)$ transformations (respectively):
\begin{eqnarray*}
{\overline{\psi}_o} \rightarrow {\overline{\psi}_o}
\exp(i\theta) 
\hspace{1.0 in}{\psi_o} \rightarrow  \exp(i\theta){\psi_o}
\\\nonumber
{\overline{\psi}_e} \rightarrow {\overline{\psi}_e}
\exp(-i\theta)	
\hspace{1.0 in}{\psi_e} \rightarrow \exp(-i\theta){\psi_e} 
\end{eqnarray*}
\begin{eqnarray*}
{\overline{\psi}_o} \rightarrow 
{\overline{\psi}_o}V_{L}^{\dagger} 
\hspace{1.0 in} {\psi_o} \rightarrow {\psi_o}
\\\nonumber
{\overline{\psi}_e} \rightarrow {\overline{\psi}_e}	
\hspace{1.0 in} {\psi_e} \rightarrow V_{L}{\psi_e} 
\end{eqnarray*}
$SU_R(2)$ is obtained by $V_L \Leftrightarrow V_R$ and $o \Leftrightarrow e$.
Here $V_L$ and $V_R$ are $SU(2)$ matrices and can be parameterized by:
$\exp(i\vec{\theta}\cdot\vec{\sigma})$ where $\sigma_i$ is a Pauli matrix
that acts on the flavor space. At $\tilde c\neq 0$, $U_A(1)$ is explicitly 
broken and the action is invariant under $SU_L(2) \times SU_R(2) \times Z_2$. 
Thus, the coupling $\tilde c$ induces the effects of the anomaly. Further 
at $m\neq 0$, it is necessary to set $V_L=V_R$ for the action to
remain invariant. Thus, with a mass term the chiral symmetry 
$SU_L(2)\times SU_R(2)$ is explicitly broken down to $SU_V(2)$. In order
to mimic QCD we need to set $\tilde c\neq0$ and $m\neq0$. Hence, our model has 
the same chiral symmetry as full QCD. Further, based on previous 
mean field strong coupling calculations \cite{Klu82}, we expect that 
the symmetry breaking pattern is also similar to full QCD. 

\section{Mapping to a Monomer - Dimer - Pion Loop - Instanton Model}
\begin{figure}[htb]
\begin{center}
\includegraphics[width=12cm]{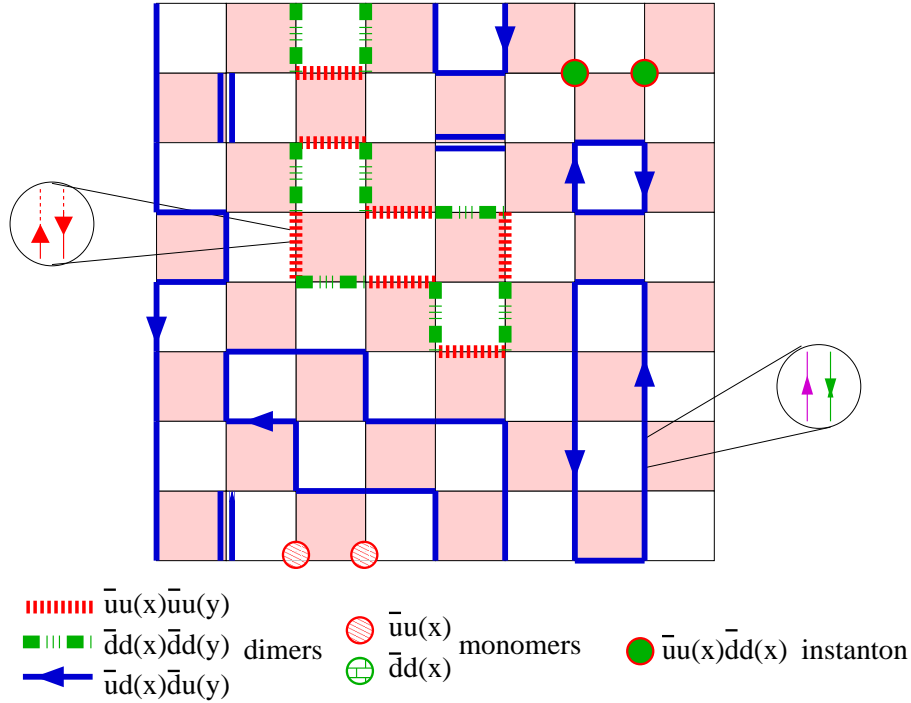}
\end{center}
\caption{\label{fig0}An example of a $2\times 2$ lattice configuration
as discussed in the text. }
\end{figure}

The partition function of our model is equivalent to that of 
a classical statistical mechanics model involving configurations made up of 
gauge invariant objects such as monomers, dimers, pion loops and 
instantons \cite{UW84}. We denote these as {\em MDPI configurations}. Note that 
a double monomer on a site breaks the $U_A(1)$ symmetry but not the 
$SU_L(2)\times SU_R(2)$ symmetry and hence is called an {\em instanton}. 
In addition to these, each configuration can contain open loops of dimers 
which terminate on monomers, closed loop of dimers and oriented closed 
pion loops made up of oriented dimers. Explicitly, the partition function 
is given by:
\begin{equation}
Z = \sum_{[I,n^d,n^u,\pi_{\mu}^d,\pi_{\mu}^u,\pi_{\mu}^1]} 
\prod_{x,\mu}
m^{n_d(x)} m^{n_u(x)} c^{I(x)}
\end{equation}
where $[I,n^d,n^u,\pi_{\mu}^d,\pi_{\mu}^u,\pi_{\mu}^1]$ denotes
a MDPI configuration. Note $I(x)$ is the number of 
instantons on a site $x$, $n_d(x)$ the number of $d$ monomers, 
$n_u(x)$ the number of $u$ monomers, $\pi_{\mu}^d$ the number of 
$d$ dimers, $\pi_{\mu}^u$ the number of $u$ dimers, and 
$\pi_{\mu}^1$ the number of oriented $\overline{u}d$ or $\overline{d}u$ 
dimers. The allowed values are: 
\begin{equation}
I(x) = 0,2, \quad n_d(x) = 0,1 \quad n_u(x) = 0,1 \quad
\pi_{\mu}^d(x) = 0,1 \quad \pi_{\mu}^u(x) = 0,1 \quad 
\pi_{\mu}^1(x) = -1,0,1
\label{eq8}
\end{equation}
Due to the Grassmann nature of the observables the following constraints 
must also be satisfied:
\begin{subequations}
\begin{eqnarray}
\sum_{\mu}\pi_{\mu}^1(x) &=& 0 \\
I(x) + \sum_{\mu}\bigg[\pi_{\mu}^u(x) + \pi_{\mu}^d(x)+ n^u(x) 
+ n^d(x)\bigg] + \sum_{\mu}\big|\pi_{\mu}^1(x)\big| &=& 2 \\
n_u(x) + \sum_{\mu}\bigg[\pi_{\mu}^u(x) - \pi_{\mu}^d(x)\bigg] 
- n_d(x) &=& 0
\end{eqnarray}
\end{subequations}
Figure \ref{fig0} gives an illustration of an MDPI configuration in
$1+1$ dimensions.

\section{Algorithm and Observables}

A Directed Path Algorithm can be constructed to update the MDPI configurations.
Our algorithm is an extension of the algorithms discussed in \cite{DA03,FJ06}.
We have three update routines:  
\begin{enumerate}
\item  A loop flip is an operation in which the orientation of a pion loop 
or a dimer loop is changed. Note that each dimer loop also contains two 
states.
\item  A loop swap which flips a dimer loop to a pion loop and vice versa.
\item  An update which creates and destroys instantons, double 
bonds, dimer loops, and pion loops.  This is similar to the directed 
loop update constructed in \cite{DA03,FJ06} and will be discussed in detail
elsewhere.
\end{enumerate}

Numerous observables can be measured with this algorithm. The simplest are
the three helicity moduli or current susceptibilities.  In particular, for 
a conserved current $J_{\mu}^i(x)$, the helicity modulus 
(current susceptibility) is defined as:  
\begin{equation}
Y^i_w = \frac{1}{d L^d}\bigg\langle 
\sum_{\mu=1}^d\bigg(\sum_x J_{\mu}^i(x)\bigg)^2\bigg\rangle
\end{equation}
where we are assuming a $L_t \times L^d$ lattice. There are three 
conserved currents in our model. They are the axial, chiral, and vector 
currents which are given by:
\begin{subequations}
\begin{eqnarray}
J_{\mu}^A(x) &=& (-1)^x\big[\pi_{\mu}^u(x)+\pi_{\mu}^d(x)+|\pi_{\mu}^1(x)|
\big] \\
J_{\mu}^C(x) &=& (-1)^x\big[\pi_{\mu}^u(x)-\pi_{\mu}^d(x)\big] \\
J_{\mu}^V(x) &=& \pi_{\mu}^1(x) \\\nonumber
\end{eqnarray}
\end{subequations}
We also can measure correlation functions of pions defined as:
\begin{subequations}
\begin{eqnarray}
G_\pi(x,y) &=& \frac{(-1)^{x+y}}{2}
\langle \overline{\psi}_x\sigma^3\psi_x \overline{\psi}_y\sigma^3\psi_y\rangle 
\\
G_\eta(x,y) &=& \frac{(-1)^{x+y}}{2}
\langle \overline{\psi}_x\psi_x \overline{\psi}_y\psi_y\rangle 
\end{eqnarray}
\end{subequations}
The corresponding susceptibilities, $\chi_\pi$ and $\chi_\eta$ are 
given by:
\begin{equation}
\chi = \frac{1}{L_t L^d} \sum_{x,y} G(x,y)
\end{equation}
The directed path algorithm allows a straightforward measurement of
$G(x,y)$ and $\chi$. Details will be given elsewhere.

\section{Results}

\begin{table}
\begin{center}
\begin{tabular}{c c c c c c c c c c} \hline\hline
&\em T &\em c &\em m & \em Algo. &\em Exact  & \em Algo. &\em Exact 
&\em Algo. &\em Exact\\\hline
& & & &$Y_w^A$ & &$Y_w^C$ & &$Y_w^V$ & \\
&1.0 &0.5 &0.0 &0.8023(9) &0.80246... &0.5763(6) &0.57721... 
&0.5771(8) &0.57721... \\ 
&1.5 &0.5 &0.0 &0.5212(7) &0.52141... &0.3275(6) &0.32790...
&0.3274(7) &0.32790...\\ 
&1.0 &1.0 &0.0 &0.7449(8) &0.74534... &0.5470(7) &0.54658...
&0.5468(8) &0.54658..\\ 
&1.5 &1.0 &0.0 &0.4967(7) &0.49720... &0.3178(5) &0.31821...
&0.3175(6) &0.31821...\\
&1.0 &1.5 &0.0 &0.6667(8) &0.66645... &0.5016(5) &0.50240...
&0.5014(6) &0.50240...\\
&1.5 &1.5 &0.0 &0.4607(6) &0.46147...&0.3029(5) &0.30325...
&0.3033(6) &0.30325...\\
&1.0 &2.0 &0.0 &0.5812(7) &0.58064... &0.4519(6) &0.45161...
&0.4521(6) &0.45161...\\
&1.5 &2.0 &0.0 &0.4199(5) &0.41927...&0.2853(5) &0.28451...
&0.2853(5) &0.28451...\\
\hline\hline
\end{tabular}
\caption{Helicity moduli for a $2\times2$ lattice as discussed in the text.}
\end{center}
\label{Tab1}
\end{table}

\begin{table}
\begin{center}
\begin{tabular}{c c c c c c c c c } \hline\hline
& & & &$\chi_\pi$ & &$\chi_\eta$ & \\
&\em T &\em c &\em m & \em Algo. &\em Exact  &\em Algo. &\em Exact \\\hline
&1.0 &0.5 &0.0 &0.3601(3) &0.359877... &0.2172(2) &0.217334... \\ 
&1.5 &0.5 &0.0 &0.2676(2) &0.267764... &0.1824(1) &0.182429...\\ 
&1.0 &1.0 &0.0 &0.4040(3) &0.403727... &0.1429(2) &0.142857...\\ 
&1.5 &1.0 &0.0 &0.2981(2) &0.298322... &0.1367(1) &0.136731...\\
&1.0 &1.5 &0.0 &0.4220(3) &0.422301... &0.0801(1) &0.080103...\\
&1.5 &1.5 &0.0 &0.3172(3) &0.317264... &0.0948(1) &0.094767...\\
&1.0 &2.0 &0.0 &0.4194(2) &0.419355... &0.0323(1) &0.032258...\\
&1.5 &2.0 &0.0 &0.3251(3) &0.324754... &0.0588(1) &0.058961...\\
\hline\hline
\end{tabular}
\caption{Susceptibilities for a $2\times2$ lattice as discussed in the text.}
\end{center}
\end{table}

We have compared the measured observables to exact values on $2\times2$ 
lattices and found excellent agreement as shown in Tables 1,2.  Below are the 
analytic expressions for the partition function, the chiral, vector, and 
axial helicity moduli and the two susceptibilities, defined in the 
previous section, for a $2\times 2$ lattice (Note that 
$c = \tilde c + m^2$):
\begin{subequations}
\begin{eqnarray}
Z(T,c,m) &=& 36T^4 + 64T^2 + 36 + c^4 + 12(1+T^2)c^2 + 8(1+T)c^2m^2  
\\\nonumber
&+& 32Tcm^2 + 16(1+T^2)m^4 + 48(1+T^3)m^2 + 32(T+T^2)m^2 
\\
2Z \times Y^C_W(T,c,m) &=& 96 + 64T^2 + 16c^2 + 112m^2 + 32m^4 + 
8c^2m^2 + 32Tcm^2 \\\nonumber &+& 32(2T+T^2)m^2
\\
2Z \times Y^V_W(T,c,m) &=& 96 + 64T^2 + 16c^2 + 64m^2
\\
Y^A_W(T,c,m) &=& Y^C_W(T,c,m) + \frac{64T^2m^2}{2Z}
\\
2Z \times \chi_\pi &=& 24T^3 + 16T^2 + 16T + 24 + 4(1+T)c^2 + 
4(3+4T+3T^2)c + 2c^3 \\\nonumber &+& 8(1+T)cm^2 + 16(1+T+T^2)m^2\\
2Z \times \chi_\eta &=& 24T^3 + 16T^2 + 16T + 24 + 4(1+T+m^2)c^2 
- 4(3-4T+3T^2)c \\\nonumber &-& 2c^3 - 8(1+T)cm^2 + 8(5+6T+5T^2)m^2 
+ 16(1+T)m^4
\end{eqnarray}
\end{subequations}
The exact analytic results are shown in Tables 1,2 along with values 
calculated from our algorithm for various $T$,$c$ and $m$.  

\section{Conclusions and Future Work}

Based on the results in the previous section, we conclude that our 
algorithm can in principle be used to study $N_f=2$ lattice QED. 
Based on previous experience we expect the algorithm to be 
efficient for a range of parameters. We plan to study the model in
$d=4$ dimensions and match our results to chiral perturbation theory.
We will assume that the lattice cutoff is of the order of $1GeV$,
and tune $T$ and $c$ so that our $F_\pi$ and $m_\eta$ are close to 
their physical values. We then plan to make connections of our data
to chiral perturbation theory as we change $m$. Ultimately, we also 
plan to compute the effects of the quark mass on pion scattering by 
measuring the appropriate two and four point correlation functions 
and extracting scattering phase shifts and lengths via Luscher's 
method as done in \cite{KO06}.

\section{Acknowledgments}

This work was done in collaboration with S.Chandrasekharan. 
It was partially supported by the DOE grant DE-FG02-05ER41368.  
The author wishes to thank Fu-Jiun Jiang for useful discussions.

\end{document}